\documentclass{ws-procs9x6}

\newcommand{\be}{\begin{equation}} 
\newcommand{\ee}{\end{equation}}
\newcommand{\ba}{\begin{eqnarray}}
\newcommand{\ea}{\end{eqnarray}}

\def\citebk#1{[\hspace{0.9mm}\raisebox{-1.85mm}[0mm][0mm]
  {\Large\cite{#1}}\hspace{-0.1mm}]}

\begin{document}

\title{Dynamics of QCD in a strong magnetic field}

\author{V.A.~Miransky$^{*}$}

\address{Department of Applied Mathematics,\\
University of Western Ontario ,\\
London, Ontario N6A 5B7, Canada \\}


\maketitle

\abstracts{QCD in a strong magnetic field yields an example of a rich, 
sophisticated and controllable dynamics.}

\section{Prologue}

On the first day of
this wonderful meeting, I decided to change the topic of my talk. 
Already during the first Session, I felt it
necessary to present something special, something connected with Arkady. 
So, I decided to talk about the dynamics of QCD in a strong magnetic
field. This talk, it seemed to me, was appropriate
indeed. The point is that the last time when I interacted with Arkady in
person was in April 1994, in the Institute for Theoretical Physics in
Santa Barbara. At that time,
we, Valery Gusynin, Igor Shovkovy, and myself, had just finished
our first work \citebk{GMS1} in the series of papers concerning the role
of
a magnetic field in dynamical symmetry breaking in 2+1 and 3+1
dimensional field theories (actually, at that time, we discussed only
the 2+1 dimensional case). So, it seemed to be appropriate to
``continue" that discussion at the Arkadyfest in Minneapolis 8 years
later.

I was lucky with this decision.
At the time of its presentation, this work had not been yet
completed.
During my talk, Arkady raised a question concerning the
dynamics of the magnetic catalysis in QCD with a large number of colors,
$N_c \to \infty$. The answer to this question, Igor Shovkovy and I
got the next day, helped to finish the work \citebk{McQCD}.

\section{Introduction}
 
Since the dynamics of QCD is extremely rich and complicated, it is
important to study this theory under external conditions which provide a
controllable dynamics. On the one hand, this allows one to understand
better the vacuum structure and Green's functions of QCD, and, on the
other hand, there can exist interesting applications of such models in
themselves. The well known examples are hot QCD 
and QCD with a large baryon density.

Studies of QCD in external electromagnetic fields had started long ago
\citebk{Kawati,SMS}. A particularly interesting case is an external magnetic
field. Using the Nambu-Jona-Lasinio (NJL) model as a low energy effective
theory for QCD, it was shown that a magnetic field enhances the
spontaneous chiral symmetry breakdown. The understanding of this
phenomenon had remained obscure until a universal role of a magnetic 
field as a catalyst of chiral symmetry breaking was established in 
Refs. \citebk{GMS1,GMS2}. The general result states that a constant
magnetic field leads to the generation of a fermion dynamical mass (i.e., 
a gap in the one-particle energy spectrum) even at the weakest attractive
interaction between fermions. For this reason, this phenomenon was called
the magnetic catalysis. The essence of the effect is the dimensional
reduction $D\to D-2$ in the dynamics of fermion pairing in a magnetic
field. In the particular case of weak coupling, this dynamics is dominated
by the lowest Landau level (LLL) which is essentially $D-2$ dimensional
\citebk{GMS1,GMS2}. 

The phenomenon of the magnetic catalysis was studied in gauge
theories, in particular, in QED \citebk{QED1,QED2} 
and in QCD \citebk{Sh,KLW}. In the recent work \citebk{KLW}, it has been
suggested that the dynamics underlying the magnetic catalysis in 
QCD is weakly coupled at sufficiently large magnetic fields. 
Here we will consider this dynamical problem rigorously, from first 
principles. In fact, we show that, at sufficiently strong magnetic 
fields, $|eB| \gg \Lambda_{QCD}^2$, there exists a consistent 
truncation of the Schwinger-Dyson (gap) equation which leads to
a reliable asymptotic expression for the quark mass $m_{q}$. 
Its explicit form reads:
\be
m_{q}^2 \simeq 2 C_{1} |e_{q}B|
\left(c_{q}\alpha_{s}\right)^{2/3}
\exp\left[-\frac{4N_{c}\pi}{\alpha_{s} (N_{c}^{2}-1)
\ln(C_{2}/c_{q}\alpha_{s})}\right],
\label{gap}
\ee
where $e_{q}$ is the electric charge of the $q$-th quark and $N_{c}$
is the number of colors. The numerical factors $C_1$ and $C_2$ equal
$1$ in the leading approximation that we use. Their value, however, can
change beyond this approximation and we can only say that they are 
of order $1$. The constant $c_{q}$ is defined as follows:
\be
c_{q} = \frac{1}{6\pi}(2N_{u}+N_{d})\left|\frac{e}{e_{q}}\right|,
\ee
where $N_{u}$ and $N_{d}$ are the numbers of up and down quark 
flavors, respectively. The total number of quark flavors is $N_{f}
=N_{u}+N_{d}$. The strong coupling $\alpha_{s}$ in the last equation is 
related to the scale $\sqrt{|eB|}$, i.e.,
\be
\frac{1}{\alpha_{s}} \simeq b\ln\frac{|eB|}{\Lambda_{QCD}^2},
\quad \mbox{ where} \quad
b=\frac{11 N_c -2 N_f}{12\pi}. 
\label{coupling}
\ee
We should note that in the leading approximation  
the energy scale $\sqrt{|eB|}$ in Eq. (\ref{coupling}) 
is fixed only up to a factor of order $1$.

The central dynamical issue underlying this dynamics is the effect of
screening of the gluon interactions in a magnetic field in the region of
momenta relevant for the chiral symmetry breaking dynamics, $m_{q}^2 \ll
|k^2| \ll |eB|$. In this region, gluons acquire a mass $M_{g}$ of order
$\sqrt{N_{f}\alpha_{s}|e_{q}B|}$. This allows to separate the dynamics 
of the magnetic catalysis from that of confinement. 

Since the background magnetic field breaks explicitly the global chiral
symmetry that interchanges the up and down quark flavors, the chiral
symmetry in this problem is $SU(N_{u})_{L}\times SU(N_{u})_{R} \times
SU(N_{d})_{L}\times SU(N_{d})_{R}\times U^{(-)}(1)_{A}$.
The $U^{(-)}(1)_{A}$ is connected with the current which is an
anomaly free linear combination 
of the $U^{(d)}(1)_{A}$ and $U^{(u)}(1)_{A}$ currents.
The generation of quark masses
breaks this symmetry spontaneously down to $SU(N_{u})_{V}\times
SU(N_{d})_{V}$ and, as a result, $N_{u}^{2}+N_{d}^{2}-1$ gapless
Nambu-Goldstone (NG) bosons occur. In Sec. \ref{NG}, we derive the
effective action for the NG bosons and calculate their decay constants and
velocities.

The present analysis is heavily based on the analysis of the magnetic
catalysis in QED done by Gusynin, Miransky, and Shovkovy \citebk{QED2}.  
A crucial difference is of course the property of asymptotic
freedom and confinement in QCD. In connection with that, our second major
result is the derivation of the low energy effective action for gluons in
QCD in a strong magnetic field. The
characteristic feature of this action is its anisotropic dynamics. In
particular, the strength of static (Coulomb like) forces along the
direction parallel to the magnetic field is much larger than that in the
transverse directions.  Also, the confinement scale in this theory is much
less than that in QCD without a magnetic field. This features imply a rich
and unusual spectrum of light glueballs in this theory.

A special and interesting case is QCD with a large number of colors, in
particular, with $N_c \to \infty$ (the 't Hooft limit). The question about
it was raised by Arkady during my talk. This theory is considered in 
Sec. \ref{infty}. 

\section{Magnetic catalysis in QCD}
\label{mag-cat}

We begin by considering the Schwinger-Dyson (gap) equation for 
the quark propagator. It has the following form:
\ba
G^{-1}(x,y) &=& S^{-1}(x,y) + 4\pi\alpha_{s} \gamma^{\mu}\int G(x,z)
\Gamma^{\nu}(z,y,u) { D}_{\nu\mu}(u,x) 
d^{4} z d^{4} u
\label{SD}     
\ea
where $S(x,y)$ and $G(x,y)$ are the bare and full fermion propagators
in an external magnetic field, ${ D}_{\nu\mu}(x,y) $ is the 
full gluon propagator and $\Gamma^{\nu}(x,y,z)$ is the full amputated
vertex function. Since the coupling $\alpha_s$ related to the scale 
$|eB|$ is small, one might think that the rainbow (ladder) approximation
is reliable in this problem. However, this is not the case.
Because of the (1+1)-dimensional form of the fermion propagator
in the LLL approximation, there are relevant higher order 
contributions \citebk{QED1,QED2}. Fortunately one can solve this problem.
First of all, an important feature of the quark-antiquark pairing dynamics 
in QCD in a strong magnetic field is that this dynamics is essentially 
abelian. This feature is provided by the form of the polarization 
operator of gluons in this theory. The point is that the dynamics 
of the quark-antiquark pairing is mainly induced 
in the region of momenta $k$ 
much less than $\sqrt{|eB|}$. This implies that the magnetic field 
yields a dynamical ultraviolet cutoff in this problem. On the other
hand, while the contribution of (electrically neutral) gluons
and ghosts in the polarization operator is proportional to
$k^2$, the fermion contribution is proportional to $|e_{q}B|$
\citebk{QED2}. As a result, the fermion contribution dominates
in the relevant region with $k^2 \ll |eB|$. 

This observation implies that there are three, dynamically very 
different, scale regions in this problem. The first one is the region 
with the energy scale above the magnetic scale $\sqrt{|eB|}$.
In that region, the dynamics is essentially the same as in QCD
without a magnetic field. In particular, the running coupling 
decreases logarithmically with increasing the energy scale there. The
second region is that with the energy scale below the magnetic scale
but much larger than the dynamical mass $m_{q}$. In this region,
the dynamics is abelian like and, therefore, the dynamics of the 
magnetic catalysis is similar to that in QED in a magnetic field. 
At last, the third region is the region with the energy scale less 
than the gap. In this region, quarks decouple and a confinement 
dynamics for gluons is realized.
 
Let us first consider the intermediate region relevant for the 
magnetic catalysis. As was indicated above, the important ingredient 
of this dynamics is a large contribution of fermions to the 
polarization operator. It is large because of an (essentially)
1+1 dimensional form of the fermion propagator in a strong magnetic
field. Its explicit form can be obtained by modifying
appropriately the expression for the polarization operator
in QED in a magnetic field \citebk{QED2}:
\begin{eqnarray}
{ P}^{AB,\mu\nu} \simeq \frac{\alpha_{s}}{6\pi}
\delta^{AB} \left(k_{\parallel}^{\mu}
k_{\parallel}^{\nu}-k_{\parallel}^{2}g_{\parallel}^{\mu\nu}\right)
\sum_{q=1}^{N_{f}}\frac{|e_{q}B|}{m^{2}_{q}},
\quad |k_{\parallel}^2| \ll m_{q}^2,
\label{Pi-IR}\\
{ P}^{AB,\mu\nu} \simeq -\frac{\alpha_{s}}{\pi}
\delta^{AB} \left(k_{\parallel}^{\mu}
k_{\parallel}^{\nu}-k_{\parallel}^{2}g_{\parallel}^{\mu\nu}\right)
\sum_{q=1}^{N_{f}}\frac{|e_{q}B|}{{k_{\parallel}^2}}, 
\quad m_{q}^2 \ll |k_{\parallel}^2|\ll |eB|,
\label{Pi-UV}
\end{eqnarray}
where $g_{\parallel}^{\mu\nu}\equiv \mbox{diag}(1,0,0,-1)$ is
the projector onto the longitudinal subspace,
and $k_{\parallel}^{\mu}\equiv g_{\parallel}^{\mu\nu} k_{\nu}$
(the magnetic field is in the $x^3$ direction).
Similarly, we introduce the orthogonal projector $g_{\perp}^{\mu\nu}\equiv
g^{\mu\nu}-g_{\parallel}^{\mu\nu}=\mbox{diag}(0,-1,-1,0)$ and
$k_{\perp}^{\mu}\equiv g_{\perp}^{\mu\nu} k_{\nu}$ that
we shall use below. Notice that quarks in a strong magnetic field
do not couple to the transverse subspace spanned by $g_{\perp}^{\mu\nu}$
and $k_{\perp}^{\mu}$. This is because in a strong magnetic field 
only the quark from the LLL matter and they couple only to the
longitudinal components of the gluon field. The latter property 
follows from the fact that spins of the LLL quarks are polarized 
along the magnetic field (see the second paper in \citebk{GMS2}).  

The expressions (\ref{Pi-IR}) and (\ref{Pi-UV}) coincide with 
those for the polarization operator 
in the $1+1$ dimensional {\it massive} QED (massive Schwinger model)
\citebk{Sch}
if the parameter $\alpha_{s} |e_{q}B|/2$ here is replaced by 
the dimensional coupling $\alpha_{1}$ of $QED_{1+1}$. As in the 
Schwinger model, Eq.~(\ref{Pi-UV}) implies that there is a massive 
resonance in the $k_{\parallel}^{\mu}k_{\parallel}^{\nu}
-k_{\parallel}^{2} g_{\parallel}^{\mu\nu}$ component of the gluon 
propagator. Its mass is
\be
M_{g}^2= \sum_{q=1}^{N_{f}}\frac{\alpha_{s}}{\pi}|e_{q}B|=
(2N_{u}+N_{d}) \frac{\alpha_{s}}{3\pi}|eB|.
\label{M_g}
\ee
This is reminiscent of the pseudo-Higgs effect in the 
(1+1)-dimensional massive QED. It is not the genuine Higgs effect 
because there is no complete screening of the color charge in the 
infrared region with $|k_{\parallel}^2|\ll m_{q}^2$. This can 
be seen clearly from Eq.~(\ref{Pi-IR}). Nevertheless, the pseudo-Higgs 
effect is manifested in creating a massive resonance and this 
resonance provides the dominant forces leading to chiral
symmetry breaking.  

Now, after the abelian like structure of the dynamics in this 
problem is established, we can use the results of the analysis in 
QED in a magnetic field \citebk{QED2} by introducing appropriate 
modifications. The main points of the analysis are: (i) the so
called improved rainbow approximation is reliable in this problem
provided a special non-local gauge is used in the analysis, 
and (ii) for a small coupling $\alpha_{s}$ 
($\alpha$ in QED), the relevant region of momenta in
this problem is $m_{q}^2 \ll |k^2| \ll |eB|$. We recall 
that in the improved rainbow approximation the vertex 
$\Gamma^{\nu}(x,y,z)$ is taken to be bare and the gluon propagator 
is taken in the one-loop approximation. Moreover, as we argued 
above, in this intermediate region of momenta, only the contribution
of quarks to the gluon polarization tensor (\ref{Pi-UV}) matters.
[It is appropriate to call this approximation the 
``strong-field-loop improved rainbow approximation". It
is an analog of the hard-dense-loop improved rainbow approximation in
QCD with a nonzero baryon density].
As to the modifications, they are purely kinematic: the overall 
coupling constant in the gap equation $\alpha$ and the dimensionless
combination $M_{\gamma}^2/|eB|$ in QED have to be replaced by 
$\alpha_s(N_{c}^2-1)/2N_{c}$ and $M_{g}^2/|e_{q} B|$, respectively. 
This leads us to the expression (\ref{gap}) for the dynamical gap.
   
After expressing the magnetic field in terms of the running coupling, 
the result for the dynamical mass takes the following convenient form:
\be
m_{q}^2 \simeq 2C_{1}\left|\frac{e_{q}}{e}\right| \Lambda_{QCD}^2 
\left(c_{q}\alpha_{s}\right)^{2/3}
\exp\left[\frac{1}{b\alpha_{s}}-\frac{4N_{c}\pi}{\alpha_{s} 
(N_{c}^{2}-1) \ln(C_{2}/c_{q}\alpha_{s})}\right].
\label{gap-vs-alpha}
\ee
As is easy to check, the dynamical mass of the $u$-quark 
is considerably larger than that of the $d$-quark. It is also
noticeable that the values of the $u$-quark dynamical mass
becomes comparable to the vacuum value 
$m^{(0)}_{dyn}\simeq 300 \mbox{ MeV}$
only when the coupling constant gets as small as $0.05$. 

\begin{figure}
\begin{center}
\epsfxsize=8.0cm
\epsffile[88 4 488 252]{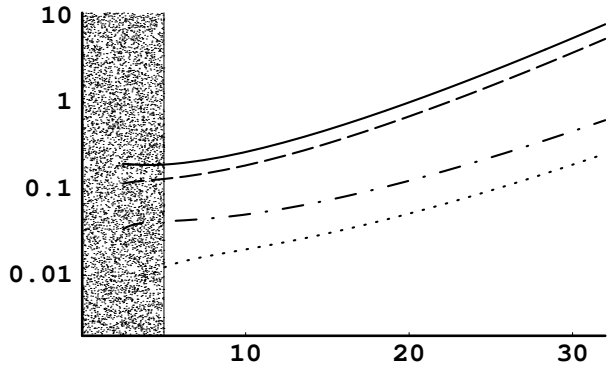}
\caption{The dynamical masses of quarks as functions of 
$\ln(|eB|/\Lambda_{QCD}^2)$ for $N_{c}=3$ and two different
values of $N_{f}=N_{u}+N_{d}$: (i) masses of $u$-quark (solid line) 
and $d$-quark (dash-dotted line) for $N_{u}=1$ and $N_{d}=2$;
(ii) masses of $u$-quark (dashed line) and $d$-quark (dotted line) 
for $N_{u}=2$ and $N_{d}=2$. The result may not be reliable in the 
weak magnetic field region (shaded) where some of the approximations 
break. The values of masses are given in units 
of $\Lambda_{QCD} = 250 \mbox{ MeV}$.}
\label{fig:rho_vs_LogB}
\end{center}
\end{figure}

Now, by trading the coupling constant for the magnetic field scale
$|eB|$, we get the dependence of the dynamical mass 
on the value of the external field. The numerical results 
are presented in  Fig. \ref{fig:rho_vs_LogB} [we used 
$C_{1} = C_{2} = 1$ in Eq. (\ref{gap-vs-alpha})]. 
As one can see in Fig. \ref{fig:rho_vs_LogB}, the value of the quark gap
in a wide window of strong magnetic fields, $\Lambda_{QCD}^{2}\ll |eB|
\lesssim (10 \mbox{ TeV})^{2}$, remains smaller than the dynamical mass of
quarks $m^{(0)}_{dyn} \simeq 300 \mbox{ MeV}$ in QCD without a magnetic field.  
In other words, the chiral condensate is partially {\it suppressed} for
those values of a magnetic field.
The explanation of this, rather unexpected, result is
actually simple. The magnetic field leads to the mass $M_g$ (\ref{M_g})  
for gluons. In a strong enough magnetic field,
this mass becomes larger than the characteristic gap
$\Lambda$ in QCD without a magnetic field ($\Lambda$, playing the role of
a gluon mass, can be estimated as a few times larger than
$\Lambda_{QCD}$). This, along with the property of the asymptotic freedom
(i.e., the fact that $\alpha_{s}$ decreases with increasing the magnetic
field), leads to the suppression of the chiral condensate.

This point also explains why our result for the gap is so different from
that in the NJL model in a magnetic field \citebk{Kawati}. Recall that, in
the NJL model, the gap logarithmically (i.e., much faster than in the
present case) grows with a magnetic field. This is the related to the
assumption that both the dimensional coupling constant $G =
g/\Lambda^2$ (with $\Lambda$ playing a role similar to that of the gluon
mass in QCD), as well as the scale $\Lambda$ do not dependent on the value
of the magnetic field. Therefore, in that model, in a strong enough
magnetic field, the value of the chiral condensate is overestimated.

The picture which emerges from this discussion is the following.
For values of a magnetic field $|eB| \lesssim \Lambda^2$ the
dynamics in QCD should be qualitatively similar to that in
the NJL model. For strong values of the field, however, it
is essentially different, as was described above. This in turn
suggests that there should exist an intermediate region of
fields where the dynamical masses of quarks decreases with
increasing the background magnetic field.

\section{Effective action of NG bosons}
\label{NG}
   
The presence of the background magnetic field breaks explicitly the global
chiral symmetry that interchanges the up and down quark flavors. This is
related to the fact that the electric charges of the two sets of quarks
are different. However, the magnetic field does not break the global
chiral symmetry of the action completely.  In particular, in the model
with the $N_{u}$ up quark flavors and the $N_{d}$ 
down quark flavors, the action is invariant under the chiral symmetry
$SU(N_{u})_{L}\times SU(N_{u})_{R} \times SU(N_{d})_{L}\times
SU(N_{d})_{R}\times U^{(-)}(1)_{A}$. 
The $U^{(-)}(1)_{A}$ is connected with the current which is an
anomaly free linear combination
of the $U^{(d)}(1)_{A}$ and $U^{(u)}(1)_{A}$ currents.
[The $U^{(-)}(1)_{A}$ symmetry is of course absent if either
$N_d$ or $N_u$ is equal to zero].

The global chiral symmetry of the action is broken spontaneously down to
the diagonal subgroup $SU(N_{u})_{V}\times SU(N_{d})_{V}$ when 
dynamical masses of quarks are generated. In agreement with the
Goldstone theorem, this leads to the
appearance of $N_{u}^{2}+N_{d}^{2}-1$ number of the NG gapless excitations
in the low-energy spectrum of QCD in a strong magnetic field. Notice
that there is also a pseudo-NG boson connected with the conventional
(anomalous) $U(1)_A$ symmetry which can be rather light in a
sufficiently strong magnetic field.
   
Now, in the chiral limit, the general structure of the low energy action
for the NG bosons could be easily established from the symmetry arguments
alone. First of all, such an action should be invariant with respect to
the space-time symmetry $SO(1,1)\times SO(2)$ which is left unbroken by
the background magnetic field [here the SO(1,1) and the SO(2) are
connected with Lorentz boosts in the $x_{0}-x_{3}$ hyperplane and
rotations in the $x_{1}-x_{2}$ plane, respectively]. Besides that, the
low-energy
action should respect the original chiral symmetry $SU(N_{u})_{L}\times
SU(N_{u})_{R} \times SU(N_{d})_{L}\times SU(N_{d})_{R}\times
U^{(-)}(1)_{A}$. 
These
requirements lead to the following general form of the action:
\ba
{ L}_{NG} &\simeq &\frac{f_{u}^{2}}{4}
\mbox{tr} \left( g_{\parallel}^{\mu\nu}
\partial_{\mu}\Sigma_{u}\partial_{\nu}\Sigma_{u}^{\dagger}
+v_{u}^{2} g_{\perp}^{\mu\nu}
\partial_{\mu}\Sigma_{u}\partial_{\nu}\Sigma_{u}^{\dagger}\right)
\nonumber\\
&+&\frac{f_{d}^{2}}{4}
\mbox{tr} \left(g_{\parallel}^{\mu\nu}
\partial_{\mu}\Sigma_{d}\partial_{\nu}\Sigma_{d}^{\dagger}
+v_{d}^{2} g_{\perp}^{\mu\nu}
\partial_{\mu}\Sigma_{d}\partial_{\nu}\Sigma_{d}^{\dagger}\right)
\nonumber\\
&+& \frac{\tilde{f}^{2}}{4}
\left(g_{\parallel}^{\mu\nu}
\partial_{\mu}\tilde{\Sigma}\partial_{\nu}
\tilde{\Sigma}^{\dagger}
+\tilde{v}^{2} g_{\perp}^{\mu\nu}
\partial_{\mu}\tilde{\Sigma}\partial_{\nu}\tilde{\Sigma}^{\dagger}\right).
\label{low-e-NG}
\ea
The unitary matrix fields $\Sigma_{u}\equiv \exp \left(
i\sum_{A=1}^{N_{u}^2-1}\lambda^{A}\pi_{u}^{A}/f_{u} \right)$,
$\Sigma_{d}\equiv \exp \left( i\sum_{A=1}^{N_{d}^2-1}
\lambda^{A}\pi_{d}^{A}/f_{d} \right)$, and 
$\tilde{\Sigma} \equiv \exp 
\left({i\sqrt{2}}\tilde{\pi}/\tilde{f} \right)$ 
describe the NG bosons in the up,
down, and $U^{(-)}(1)_{A}$
sectors of the original theory. The decay constants 
$f_{u}, f_{d}, \tilde{f}$ and transverse velocities $v_{u}, v_{d}, 
\tilde{v}$
can be calculated by using the standard field theory formalism. 
The results for  
the $N_{u}^{2}+N_{d}^{2}-2$ NG bosons in the up and down sectors,
assigned to the adjoint representation of 
the $SU(N_{u})_{V}\times SU(N_{d})_{V}$ symmetry, are \citebk{McQCD}
\ba
f_{u}^{2} &=& \frac{N_c}{6\pi^{2}}|eB|,\\ 
f_{d}^{2} &=& \frac{N_c}{12\pi^{2}}|eB|,
\quad v_{q}=0.
\ea
The remarkable fact is that the decay constants are nonzero
even in the limit when the dynamical masses of quarks 
approach zero. The reason of that is the $1+1$ dimensional
character of this dynamics.
A similar situation takes place in color
superconductivity: in that case the
$1+1$ dimensional character of the dynamics is provided by
the Fermi surface. 

Notice that the transverse velocities of the NG bosons are equal to zero.
This is also a consequence of the $1+1$ dimensional structure of the quark
propagator in the LLL approximation.  The point is that quarks can move in
the transverse directions only by hopping to higher Landau levels. Taking
into account higher Landau levels would lead to nonzero velocities
suppressed by powers of $|m_{q}|^{2}/|eB|$. In fact, the explicit form of
the velocities was derived in the weakly coupled NJL model in an
external magnetic field [see Eq.~(65) in the second paper of 
Ref.~\citebk{GMS2}].
It is
\be
v_{u,d}^{2} \sim \frac{|m_{u,d}|^{2}}{|eB|}
\ln\frac{|eB|}{|m_{u,d}|^{2}} \ll 1.
\ee
A similar expression should take place also for the transverse 
velocities of the NG bosons in QCD.

The decay constant $\tilde{f}$ of the singlet NG boson connected with the
spontaneous breakdown of the $U^{(-)}(1)_{A}$ is \citebk{McQCD}
\be 
\tilde{f}^{2} = \frac{(N_{d} f_{u} + 
N_{u} f_{d})^2}{N^{2}_{f}}=
\frac{(\sqrt{2}N_{d} + N_{u})^{2}N_c}{12\pi^{2}N^{2}_{f}}|eB|.
\ee
Its transverse velocity is of course zero in the LLL approximation.

\section{Anisotropic confinement of gluons}
\label{conf}

Let us now turn to the infrared region with $|k|\lesssim m_d$, where all
quarks decouple (notice that we take here the smaller mass of $d$
quarks). In that region, a pure gluodynamics realizes. However, its
dynamics is quite unusual. The point is that although gluons are
electrically neutral, their dynamics is strongly influenced by an external
magnetic field, as one can see from expression (\ref{Pi-IR}) for their
polarization operator. In a more formal language, while quarks decouple
and do not contribute into the equations of the renormalization group in
that infrared region, their dynamics strongly influence the boundary
(matching) conditions for those equations at $k \sim m_d$.
A conventional way to describe this dynamics is the method of the low
energy effective action. This low effective action was derived
in Ref. \citebk{McQCD}. Here we will discuss its main properties.

The low energy effective action is relevant for momenta $|k| \lesssim
m_d$. Notice the following important feature of the action: its 
``bare" coupling constant $g$, related to the 
scale $m_d$, coincides with the value of the vacuum QCD coupling related
to the scale $\sqrt{|eB|}$ (and {\it not} to the scale $m_d$). This is
because
$g$ is determined from the matching condition at $|k| \sim m_d$, the lower
border of the intermediate region $m_d \lesssim |k| \lesssim \sqrt{|eB|}$,
where,
because of the pseudo-Higgs effect, the running of the coupling is
essentially frozen. Therefore the ``bare" coupling $g$ indeed coincides
with the value of the vacuum QCD coupling related to 
the scale $\sqrt{|eB|}$: $g = g_s$. Since this value is much less that
that of the vacuum QCD coupling
related to the scale $m_d$, this implies that the confinement scale
$\lambda_{QCD}$ of the action should be much less
than $\Lambda_{QCD}$ in QCD without a magnetic field.

Actually, this consideration somewhat simplifies the real
situation. Since the
LLL quarks couple to the longitudinal components of the
polarization operator, only the effective coupling connected
with longitudinal gluons is frozen. For transverse gluons, there
should be a logarithmic running
of their effective coupling. It is clear, however, that this
running should be quite different from that in the vacuum QCD.
The point is that the time like gluons are now massive and 
their contribution in the running in the intermediate region
is severely reduced. On the other hand, 
because of their negative norm, just the time like gluons
are the major players in producing the antiscreening running
in QCD (at least in covariant gauges). 
Since now they effectively decouple,
the running of the effective coupling for the transverse gluons
should slow down. It is even not inconceivable that the
antiscreening running can be transformed into a screening one.  
In any case, one should expect that the value of the transverse
coupling related to the matching scale $m_d$ will be also essentially
reduced in comparison with that in the vacuum QCD. Since the
consideration in this section is rather qualitative, 
we adopt  the simplest scenario with the value of the transverse coupling
at the matching scale $m_d$ also
coinciding with $g_s$.  

The interaction potential
between two static quarks in this theory at ''short" distances
$r \sim m_{d}^{-1}$ reads:
\be
V(x,y,z) \simeq \frac{g_{s}^{2}}
{4\pi\sqrt{z^2+\epsilon (x^{2}+y^{2})}},
\label{potent}
\ee
where the dielectric constant $\epsilon = 1 + \frac{\alpha_{s}}{6\pi}
\sum_{q=1}^{N_{f}}|e_{q}B|/m^{2}_{q}$ is very large.
Because of the dielectric constant, this Coulomb like interaction is
anisotropic in space: it is suppressed by a factor of $\sqrt{\epsilon}$ in
the transverse directions compared to the interaction in the direction of
the magnetic field. 

The potential (\ref{potent}) corresponds to the
classical, tree, approximation which is good only in the region of
distances much smaller than the confinement radius $r_{QCD} \sim
\lambda^{-1}_{QCD}$. Deviations from this interaction are determined 
by loop corrections. The analysis of the loop
expansion leads to the following estimate of the new confinement scale
$\lambda_{QCD}$ in QCD in a strong magnetic field:
\be
\lambda_{QCD} \simeq m_d
\left(\frac{\Lambda_{QCD}}{\sqrt{|eB|}}\right)^{b/b_{0}},
\label{lambda}
\ee
where $b=(11 N_c -2 N_f)/12\pi$ and  $b_{0}=11 N_c/12\pi$. Therefore,
in a strong magnetic field,
$\lambda_{QCD}$ is much less than $\Lambda_{QCD}$.

The hierarchy $\lambda_{QCD} \ll \Lambda_{QCD}$ is intimately connected
with a somewhat puzzling point that the pairing dynamics decouples
from the confinement dynamics
despite it produces quark masses of order
$\Lambda_{QCD}$ or less [for a magnetic field all the way up to the order
of $(10 \mbox{ TeV})^2$]. The point is that these masses are heavy in
units
of the new confinement scale $\lambda_{QCD}$ and the pairing dynamics
is indeed weakly coupled.

\section{Arkady question: QCD with a large number of colors }
\label{infty}

I did not discuss the case of QCD with a
large
number of colors in my talk. The question about it was raised by Arkady 
during the talk (he
actually asked not myself, the speaker, but my coauthor Igor
Shovkovy who was sitting next to him). Igor and myself got the
answer the next day. 

Just a look at expression (\ref{M_g})
for the gluon mass is enough to recognize that the dynamics in this limit
is very different from that considered in the previous sections. Indeed,
as is well known, the strong coupling constant $\alpha_s$ is proportional
to $1/N_c$ in this limit. More precisely, it rescales as
$\alpha_s = \frac{\tilde{\alpha}_s}{N_c}$,
where the new coupling constant $\tilde{\alpha}_s$ remains finite
as $N_c \to \infty$ ('t Hooft limit). 
Then, expression (\ref{M_g}) implies that
the gluon mass goes to zero in this limit. This in turn implies
that the appropriate approximation in this limit is not
the improved rainbow approximation but the rainbow approximation
itself, when {\it both} the vertex and the gluon propagator in
the SD equation (\ref{SD}) are taken to be bare. This leads to
the following expression for the dynamical mass of quarks
\citebk{McQCD}:
\be
m_{q}^2 = C |e_{q}B|
\exp\left[-{\pi}
\left(\frac{\pi N_c}{(N_{c}^2-1)\alpha_s}\right)^{1/2}
\right],
\label{m_q}
\ee
where the constant $C$ is of order one. The confinement scale
$\lambda_{QCD}$ is close to $\Lambda_{QCD}$ in this case \citebk{McQCD}.

It is natural to ask how large $N_c$ should be before the expression
(\ref{m_q}) becomes reliable. One can show \citebk{McQCD}
that the threshold value
$N^{thr}_c$ grows rapidly with the
magnetic field [$N^{thr}_c \gtrsim 100$ for 
$|eB| \gtrsim (1\mbox{GeV})^2$]. 
Expression (\ref{m_q}) for the quark mass
is reliable for the values of
$N_c$ of the order of $N_{c}^{thr}$ or larger.
Decreasing $N_c$ below $N_{c}^{thr}$, one comes to
expression (\ref{gap}).

\section{Conclusion}

QCD in a strong magnetic field yields an example of a rich, sophisticated
and (that is very important) controllable dynamics.

\section*{Acknowledgments}
I am grateful to the organizers of this Symposium for their warm
hospitality. I acknowledge support from the Natural Sciences and
Engineering Research Council of Canada.

\end{document}